\documentclass[reprint,aps,pre,tightlines,groupedaddress]{revtex4-2}
\usepackage{graphicx}
\usepackage{dcolumn}
\usepackage{bm}
\usepackage{amsmath,amssymb}
\usepackage{amsbsy}
\usepackage{eucal}
\usepackage{hyperref}
\usepackage{varioref}
\usepackage{float}
\numberwithin{equation}{section}
\usepackage[compact]{titlesec}  
\titlespacing{\section}{10pt}{10pt}{10pt}
\renewcommand{\thesection}{\arabic{section}}
\renewcommand{\thesubsection}{\thesection.\arabic{subsection}}
\renewcommand{\thesubsubsection}{\thesubsection.\arabic{subsubsection}}

\begin{document}

\title{Large Deviation Theory for Bose Gas of Photons and Planck's Oscillators}

\author{D. P. Shinde}
\affiliation{168-A, Panchsheel Colony, Pachgaon, Kolhapur, State-Maharashtra, India - 416013.}
\email{dpshinde.theory@gmail.com}

\date{\today}

\begin{abstract}
We utilize large deviation theorems to analyze the distributions of a Bose gas of photons and Planck's identical linear oscillators. By applying the Boltzmann-Sanov and Cramér-Chernoff theorems, we calculate the large deviation probabilities, entropies, and rate functions for the spatial and energy distributions of both photons and Planck's oscillators. Our study reproduces the results of Bose and Planck within the framework of large deviation theory.
\end{abstract}


\maketitle

\section{Introduction}
Modern abstract mathematical analysis methods \cite{Kolmogorov54,Petrov95}  are essential for determining the asymptotic forms of probabilistic limit theorems and the laws of large numbers concerning sums of independent and identically distributed (i.i.d.) random variables. The strong and weak laws of large numbers establish criteria for the convergence of probability measures for these random variables \cite{Kolmogorov54,Petrov95}. These methods are also crucial for calculating the distributions of various physical quantities in classical and quantum mechanical systems \cite{LandauV5,Schroedinger13,Haar95,Pathria11,Greiner95}. Physical quantities, such as internal energies and the number of systems, can be modeled as sets of i.i.d. random variables. Under generalized conditions, the probability distributions of internal energies or their sums are governed by the central limit theorem (CLT), which illustrates the Laplace-Gauss law of probability \cite{LandauV5,Schroedinger13,Haar95,Pathria11,Greiner95}. This law explains how the sums of random variables tend to deviate from their mean values, with typical deviations being on the order of the mean itself. These deviations are commonly referred to as normal deviations \cite{LandauV5,Schroedinger13,Haar95,Pathria11,Greiner95}.

Many studies \cite{Feller43,Linnik61,Sethuraman64,Petrov65,Varadhan66,Hoeffding67,Efron68,Ventsel70,Donsker75a,Donsker75b,Gartner77,Nagaev79,Ellis84,Bryc92} have shown that deviations from the normal scale of the central limit theorem referred to as large deviations, are crucial for advancing mathematical probability theory and its applications across various scientific fields. In particular, the weak law of large numbers provides a framework for examining large deviations \cite{Feller43,Linnik61,Sethuraman64,Petrov65,Varadhan66,Hoeffding67,Efron68,Ventsel70,Donsker75a,Donsker75b,Gartner77,Nagaev79,Ellis84,Bryc92}. Boltzmann \cite{Boltzmann77} was a pioneer in calculating large deviation probabilities for random variables in thermodynamic systems, establishing a foundation for future research in this area. Throughout the twentieth century, numerous studies expanded on the weak law, thoroughly exploring large deviation probabilities and their decay rates using straightforward mathematical formulations \cite{Feller43,Linnik61,Sethuraman64,Petrov65,Varadhan66,Hoeffding67,Efron68,Ventsel70,Donsker75a,Donsker75b,Gartner77,Nagaev79,Ellis84,Bryc92} and abstract probability analysis \cite{Varadhan84,Hollander00,Ellis06}. In examining physical systems, these abstract mathematical approaches consider level-1 and level-2 microscopic sums for random quantities, such as the spatial arrangements and energies of systems, as discussed in the Boltzmann-Sanov \cite{Boltzmann77,Sanov61} and Cramér-Chernoff theorems \cite{Cramer38,Chernoff52,Ellis06}. Moreover, the abstract probability measures described in the weak law of large numbers yield large deviation functions and entropies for both level-1 and level-2 microscopic sums \cite{Ellis06}. 
Furthermore, Boltzmann-Sanov's theorem and Cramér-Chernoff's theorem are related through the contraction principle \cite{Varadhan84,Hollander00,Ellis06}. 

In classical statistical mechanics \cite{Landford71,Oono89,Derrida98,Derrida02,Touchette09}, large deviation theorems are employed to analyze the spatial and energy distributions of distinguishable and indistinguishable systems of the classical canonical ensemble . In this context \cite{Shinde21}, the total internal energy and the total number of systems in an ensemble are treated as macroscopic i.i.d. random variables. These quantities are calculated as the sums of the internal energies and the number of identical and independent systems across infinitesimal partitions of the ensemble's state space \cite{Shinde21}. The possible arrangements of spatial distributions and energy levels of the systems across these infinitesimal partitions must comply with the constraints regarding the total number of systems and the total internal energy of the ensemble \cite{Shinde21}. Specifically, empirical frequency is the ratio of the number of systems found in a given partition to the total number of systems in the ensemble. The weak law of large numbers establishes the asymptotic behavior of probabilities concerning large deviations between empirical frequencies and relative volumes of the partitions of a state space \cite{Born49,Shinde21}. Notable large deviation theorems, such as those proposed by Boltzmann and Sanov \cite{Boltzmann77,Sanov61,Ellis06}, offer insights into the probability distributions related to the spatial arrangements of the systems. Importantly, the ensemble is isolated from its reservoir to accurately capture the spatial distribution of the systems, preventing any exchange of energy or systems between the ensemble and the reservoir \cite{Shinde21}. Boltzmann-Sanov relative entropy is directly related to the natural logarithm of the spatial distribution. In a state of thermodynamic or statistical equilibrium, identified as the most frequent arrangements of systems, the Boltzmann-Sanov relative entropy (the large deviation function) reaches either its maximum (minimum) value \cite{Varadhan84,Hollander00,Ellis06,Kullback51,Cover06,Shinde21}.

The interaction of an ensemble with the thermal reservoir allows for the study of the probability distribution of internal energies of various systems \cite{Ellis06,Shinde21}. The Cramér-Chernoff theorem \cite{Cramer38,Chernoff52,Ellis06} addresses large deviation probabilities related to these internal energies of systems and the ensemble as a whole. The mathematical properties of the moment-generating function (MGF), also known as the partition function of an ensemble are utilized to investigate large deviations in internal energy \cite{Ellis06,Shinde21}. By combining the MGF with the Bienayme-Chebyshev-Markov (BCM) inequality, we can derive the optimal values for the Cramer-Chernoff large deviation upper bound related to energy variables \cite{Ellis06,Shinde21}.
In this scenario, the Boltzmann-Gibbs canonical entropy is proportional to the natural logarithm of the probability distribution of energy. At thermodynamic equilibrium, the Boltzmann-Gibbs canonical entropy reaches its maximum value \cite{Ellis06,Shinde21}. Additionally, the Cramer-Chernoff large deviation function attains its minimum value \cite{Ellis06,Shinde21}. Under subsidiary conditions regarding the total number of systems and their total energy, Lagrange multipliers are introduced to address the equilibrium condition. With these constraints, the variations in Boltzmann-Sanov relative entropy and Boltzmann-Gibbs canonical entropy are eliminated \cite{Ellis06,Shinde21}. The thermodynamic equilibrium condition provides the mean values of the internal energies and entropies for a given macrostate \cite{Ellis06,Shinde21}. Furthermore, these results \cite{Shinde21} regarding large deviations are linked to thermodynamic variables, leading to a broader understanding of the generalized second law of thermodynamics \cite{Prigogine52}.

Both Bose \cite{Bose24,Bose76,Wali} and Planck \cite{Planck91,Planck00,Planck01} employed Boltzmann's \cite{Boltzmann77} theory of complexions to calculate the spatial distributions of systems. However, neither study \cite{Bose24,Bose76,Wali,Planck91,Planck00,Planck01} utilized the BCM probability inequality concerning the energy variables of the systems. The Lagrange multipliers, arising from subsidiary conditions on the total number of particles and energy, were directly incorporated into the equilibrium analysis \cite{Bose24,Bose76,Wali,Planck91,Planck00,Planck01}.

This paper explores large deviations in the distributions of random variables in Bose and Planck systems. To achieve this, we apply the large deviation theorems to the distributions of a gas of photons or light quanta \cite{Bose24,Bose76,Wali} and Planck’s identical linear oscillators \cite{Planck91,Planck00,Planck01}. We begin by examining Bose’s significant contributions, in which he developed statistics for the gas of photons using principles of quantum mechanics. Next, we examine the distributions of Planck's identical linear oscillators. For both analyses, we adopt notations similar to those used by Bose and Planck, ultimately determining the large deviation probabilities, entropy, and rate functions for the spatial and energy distributions of photons and Planck’s oscillators.

\section{Case of Photons}

Let us examine Bose's significant research on the statistics of light quanta, or photons \cite{Bose24,Bose76,Wali}. By applying quantum principles, Bose \cite{Bose24,Bose76,Wali} reproduced Planck's derivations concerning black-body radiations. His work \cite{Bose24,Bose76,Wali} analyzed the statistical distribution related to various cells containing a certain number of indistinguishable photons.

Let $p^{s}_{0},p^{s}_{1},....,p^{s}_{r},...$ represent the cells containing $r$ number of $s$-type of quanta within the infinitesimal frequency range $\nu^{s}$ and $\nu^{s}+d\nu^{s}$ \cite{Bose24,Bose76,Wali}. The indices $r$ and $s$ vary from zero to infinity. The subsidiary constraints regarding the number of cells $A^{s}$, the number of photons $N^{s}$, and their energy $E^{s}$ in the range $d\nu^{s}$ are expressed as $A^{s}=\sum_{r=0}^{\infty} p^{s}_{r}$, $N^{s}=\sum_{r=0}^{\infty}r p^{s}_{r}$, and $E^{s}= N^{s}h\nu^{s} =\sum_{r=0}^{\infty}r p^{s}_{r}h\nu^{s}$ \cite{Bose24,Bose76,Wali}. It is important to note that the constraint on $N^{s}$ is not necessary when dealing with photons \cite{Bose24,Bose76,Wali}. We observe that the spatial distribution $W^{s}_{1}$ of arrangements of the cells $p^{s}_{0},p^{s}_{1},....,p^{s}_{r},...$, follows the Boltzmann-Sanov multinomial theorem. Consequently, the spatial distribution $\mathbb{W}_{1}$ of the entire system is also represented \cite{Bose24,Bose76,Wali}.
\begin{eqnarray}
&& W^{s}_{1}(p^{s}_{0},p^{s}_{1},....,p^{s}_{r},..)= \frac{A^{s}!}{\prod \limits^{\infty}_{r=0} p^{s}_{r}!},\nonumber \\ 
&& \mathbb{W}_{1} = \prod \limits^{\infty}_{s=0} W^{s} =\prod \limits^{\infty}_{s=0} \frac{A^{s}!}{\prod \limits^{\infty}_{r=0} p^{s}_{r}!}.
\label{Eq1}
\end{eqnarray}
In classical statistical mechanics, large deviation theory assesses the discrepancy between empirical frequencies and the probabilities associated with filling or the relative volumes of cells in phase space \cite{Ellis06}. This discrepancy becomes asymptotically very small when the system is at thermodynamic equilibrium \cite{Ellis06}. In quantum statistical mechanics, large deviation theory can analyze differences in empirical frequencies based on occupation numbers and the weights assigned to different quantum cells according to their degeneracy. Additionally, the principles of quantum mechanics dictate that all non-degenerate cells or states are assigned equal weights when the system is at statistical equilibrium \cite{LandauV5,Schroedinger13,Haar95,Pathria11,Greiner95}.

In the case of photons, all cells have equal weights \cite{LandauV5,Schroedinger13,Haar95,Pathria11,Greiner95}. In his second paper \cite{Wali}, Bose examined the distribution of material particles $(N=\sum_{i}n_{i})$ and derived the thermodynamic probability equation, $\frac{N!g_{1}^{n_{1}}g_{2}^{n_{2}}...}{n_{1}!n_{2}!...}$, where $g$ represents the probability of a particle occupying a single cell. To study large deviations in systems other than photons, we consider the weights $\omega^{s}_{r}$ of the cells. This approach allows us to obtain the spatial distributions $W^{s}_{2}$ and $\mathbb{W}_{2}$.
\begin{eqnarray}
&& W^{s}_{2}(p^{s}_{0},p^{s}_{1},....,p^{s}_{r},..)= A^{s}!\prod \limits^{\infty}_{r=0}\frac{(\omega^{s}_{r})^{ p^{s}_{r}}}{ p^{s}_{r}!}, \nonumber \\ 
&& \mathbb{W}_{2}=\prod \limits^{\infty}_{s=0} W^{s} =\prod \limits^{\infty}_{s=0} A^{s}!\prod \limits^{\infty}_{r=0}\frac{(\omega^{s}_{r})^{ p^{s}_{r}}}{ p^{s}_{r}!}.
\label{Eq2}
\end{eqnarray}
First, we take the natural logarithm of both sides of Eqns.\eqref{Eq1} and \eqref{Eq2}. Next, we apply Stirling’s approximation to the large numbers for the factorial terms $A^{s}!$ and $p^{s}_{r}!$, given that $\log (A^{s}!)= A^{s}\log A^{s}-A^{s}+ \mathcal{O}(\log A^{s})$. Finally, by dividing both sides by $A^{s}$, we arrive at 
\begin{eqnarray}
&& \frac{\log W^{s}_{1}}{A^{s}}=\sum_{r=0}^{\infty} \frac{p^{s}_{r}}{A^{s}}\log\left(\frac{A^{s}}{p^{s}_{r}}\right)+ \mathcal{O}(\frac{\log A^{s}}{A^{s}}), \nonumber \\
&&\frac{\log W^{s}_{2}}{A^{s}}=\sum_{r=0}^{\infty} \frac{p^{s}_{r}}{A^{s}}\left[\log \omega^{s}_{r}-\log\left(\frac{A^{s}}{p^{s}_{r}}\right)\right]+ \mathcal{O}(\frac{\log A^{s}}{A^{s}}),
\label{Eq3} 
\end{eqnarray}
where, we have used $A^{s}=\sum_{r=0}^{\infty}p^{s}_{r}$. 
As the limit $A^{s}\to\infty$, the order term $\mathcal{O}(.)$ converges to zero \cite{Varadhan84,Hollander00,Ellis06}. With this limit, the sums on the right-hand sides of the upper and lower lines of \eqref{Eq3} are referred to as large deviation functions \cite{Varadhan84,Hollander00,Ellis06,Kobayashi11}, denoted $I_{1}^{s}=-\sum_{r=0}^{\infty} \frac{p^{s}_{r}}{A^{s}}\log\left(\frac{A^{s}}{p^{s}_{r}}\right)$ and $I_{2}^{s}=-\sum_{r=0}^{\infty} \frac{p^{s}_{r}}{A^{s}}\left[\log \omega^{s}_{r}-\log\left(\frac{A^{s}}{p^{s}_{r}}\right)\right]$, respectively. If all weights $\omega^{s}_{r}$s are equal to one, then $I_{2}^{s}=I_{1}^{s}$. Furthermore, $I_{2}^{s} = 0 $ when $\omega^{s}_{r} = p^{s}_{r}/A^{s}$. Additionally, Boltzmann's relative entropies are defined as $S_{1}^{s}= - k I_{1}^{s}$ and $S_{2}^{s} = -k I_{2}^{s}$. From Eqn \eqref{Eq3}, we can express the exponential decay of the probability using the large deviation functions  $W_{1}^{s}\approx e^{-A^{s}I_{1}}$ and $ W_{2}^{s} \approx e^{-A^{s}I_{2}}$. It is important to note that Boltzmann-Sanov relative entropy is proportional to the natural logarithm of the spatial probability distribution \cite{Varadhan84,Hollander00,Ellis06,Kobayashi11}. Ultimately, by setting $\mathbb{W}=\prod \limits^{\infty}_{s=0} W^{s}$, we can calculate the Boltzmann-Sanov relative entropies $\mathbb{S}_{1}$ and $\mathbb{S}_{2}$ for the entire system.
\begin{eqnarray}
\begin{split}
\mathbb{S}_{1} = k \log \mathbb{W}_{1} & = k \sum_{s=0}^{\infty} \left(\sum_{r=0}^{\infty}p^{s}_{r} \log \frac{A^{s}}{p^{s}_{r}}+ \mathcal{O}(\log A^{s})\right).\\ 
\mathbb{S}_{2} = k \log \mathbb{W}_{2} & = k \sum_{s=0}^{\infty} \left(\sum_{r=0}^{\infty}p^{s}_{r}(\log \omega^{s}_{r}-\log \frac{A^{s}}{p^{s}_{r}}) + \mathcal{O}(.)\right).\\
\label{Eq4}
\end{split}
\end{eqnarray}
Here, $k$ is Boltzmann constant. 

Let us now explore the Cramér-Chernoff theorem as it relates to the distributions of energies of photons. We consider the energies $0,h\nu^{s},2h\nu^{s},....,rh\nu^{s},....$, which are integral multiples of $h\nu$, as non-negative random variables \cite{khinchin13}. The Cramér-Chernoff theorem establishes an upper bound on the probability of the exponential random variable $e^{-\tilde{\beta}E^{s}}$, where $\tilde{\beta} = 1/kT>0$ \cite{Varadhan84,Hollander00,Ellis06,Kobayashi11}. This theorem employs the BCM probability inequality regarding the total energy of the system and a specific threshold value of energy \cite{Varadhan84,Hollander00,Ellis06,Kobayashi11}. In particular, the BCM probability inequality considers the moment-generating function, also known as the partition function of a photon, along with the energy threshold $rA^{s}h \nu^{s}$. The subsidiary condition on the total energy suggests this threshold value \cite{Varadhan84,Hollander00,Ellis06,Kobayashi11}. Within this framework, the large deviation upper bound for positive temperature is expressed as 
\begin{equation}
P^{s}(e^{-\tilde{\beta} E^{s}} \geq e^{-\tilde{\beta} r A^{s}h\nu^{s}}) \leq \frac{\mathbb{E}_{x}(e^{-\tilde{\beta} E^{s}})}{e^{-\tilde{\beta} r A^{s}h\nu^{s}}}, 
\label{Eq5}
\end{equation}
where $\mathbb{E}_{x}$ represents the mathematical expectation. Finally, we extend these calculations to encompass the entire system of photons.
\begin{equation}
\mathcal{P} = \prod \limits^{\infty}_{s=0} P^{s} \leq \prod \limits^{\infty}_{s=0} \frac{\mathbb{E}_{x}(e^{-\tilde{\beta} E^{s}})}{e^{-\tilde{\beta} r A^{s}h\nu^{s}}} = \prod \limits^{\infty}_{s=0} \frac{(Z^{s})^{A^{s}}}{e^{-\tilde{\beta} r A^{s}h\nu^{s}}},  
\label{Eq6}
\end{equation}
where $Z^{s}(\tilde{\beta}) = \mathbb{E}_{x}(e^{-\tilde{\beta} E^{s}})$ is the partition function \cite{khinchin13}.
Let's take the natural logarithm of both sides of Eq. \eqref{Eq6}. This gives us  
\begin{eqnarray}
&& \log \mathcal{P} \leq \sum_{s=0}^{\infty}\left[\tilde{\beta} r A^{s}h\nu^{s}+ A^{s}\log Z^{s} \right], \nonumber \\
&& \log \mathcal{P} \leq \sum_{s=0}^{\infty} \sum_{r=0}^{\infty}\left[\tilde{\beta} r p^{s}_{r} h\nu^{s}+ p^{s}_{r}\log Z^{s} \right]
\label{Eq7}.
\end{eqnarray}
Once again, we have utilized the relation $A^{s}=\sum_{r=0}^{\infty}p^{s}_{r}$. Next, if we divide both sides of Eq. \eqref{Eq7} by $A^{s}$ and then consider the limit as $A^{s} \to \infty$, we find that the large deviation  function is given by $\mathcal{I}^{s}=\max_{\tilde{\beta}}\left[-\tilde{\beta}r h\nu^{s}- \log Z^{s}\right]$. Additionally, the Boltzmann-Gibbs canonical entropy is equal to $k$ times the natural logarithm of $\mathcal{P}$.
Therefore, from Eq. \eqref{Eq7}, we can express both the canonical entropy $\mathcal{S}$ and the large deviation function $\eta$ for the entire system.
\begin{eqnarray}
\begin{split}
& \mathcal{S}_{}=\min_{\tilde{\beta}}\left(k \tilde{\beta} \sum_{s=0}^{\infty} \sum^{\infty}_{r=0} r p^{s}_{r}h\nu_{s} + k \sum_{s=0}^{\infty} \sum^{\infty}_{r=0}p^{s}_{r}\log Z^{s}\right). \\
& \eta_{}=-\frac{\mathcal{S}_{}}{k}= \max_{\tilde{\beta}}\left(-\tilde{\beta} \sum_{s=0}^{\infty} \sum^{\infty}_{r=0} r p^{s}_{r}h\nu_{s} - \sum_{s=0}^{\infty} \sum^{\infty}_{r=0}p^{s}_{r}\log Z^{s}\right).
\label{Eq8}. 
\end{split}
\end{eqnarray}
The thermodynamic equilibrium condition states that the changes in the entropies $\mathbb{S}_{1}$ or $\mathbb{S}_{2}$ and $\mathcal{S}$ must all equal zero \cite{Ellis06}, which implies $\delta \mathbb{S}_{1}-\delta\mathcal{S}_{} = 0 $ and $\delta \mathbb{S}_{2}-\delta\mathcal{S}=0$.. By taking the $\delta$ variations of both sides of Eqs. \eqref{Eq4} and \eqref{Eq8} with respect to $p^{s}_{r}$ and setting them to zero, we establish the equilibrium probability.
\begin{eqnarray}
\begin{split}
&\frac{p^{s}_{r}}{A^{s}}=\frac{e^{-\tilde{\beta} r h\nu^{s}}}{Z^{s}(\tilde{\beta})}.\\
&\frac{p^{s}_{r}}{A^{s}}=\frac{\omega^{s}_{r}e^{-\tilde{\beta} r h\nu^{s}}}{Z^{s}(\tilde{\beta})}. 
\end{split}
\label{Eq9} 
\end{eqnarray}
If all $\omega^{s}_{r}=1$, the lower line of Eq. \eqref{Eq9} is equal to the upper line. This equation represents the probability that the energy random variable takes on the value $r h\nu^{s}$ \cite{khinchin13}. The upper line of Eq.\eqref{Eq9} illustrates the Bose result \cite{Bose24,Bose76,Wali}. By substituting the expression for $p^{s}_{r}$ from Eq.\eqref{Eq9} into the subsidiary condition $\sum^{\infty}_{r=0} p^{s}_{r} = A^{s}$, we derive the partition function $Z^{s}(\tilde{\beta})=\sum^{\infty}_{r=0} e^{-\tilde{\beta} r h\nu^{s}}= (1-e^{-\tilde{\beta} h\nu^{s}})^{-1}$ of a photon. Additionally, the ratio of $A^{s}/Z^{s}$ is equal to $B^{s}$, as stated in Bose's paper \cite{Bose24,Bose76,Wali}. Furthermore, applying Eq. \eqref{Eq9} to other subsidiary conditions regarding $N^{s}$ and $E^{s}$ provides, respectively, the mean occupation number and mean energy of the photons per cell within the range $d\nu^{s}$ \cite{Bose24,Bose76,Wali}.
\begin{eqnarray}
\begin{split}
& N^{s}=\sum_{r=0}^{\infty}r p^{s}_{r}= A^{s}\frac{ \sum_{r=0}^{\infty}r e^{-\tilde{\beta} r h\nu^{s}}}{\sum^{\infty}_{r=0} e^{-\tilde{\beta} r h\nu^{s}}},\\ 
&\frac{N^{s}}{A^{s}}=\frac{1}{e^{\tilde{\beta} h\nu^{s} }-1}.
\end{split}
\label{Eq10}
\end{eqnarray}
\begin{eqnarray}
\begin{split}
& E^{s}=\sum_{r=0}^{\infty}r p^{s}_{r}h\nu^{s} = A^{s}\frac{\sum_{r=0}^{\infty}r h\nu^{s} e^{-\tilde{\beta} r h\nu^{s}}}{\sum^{\infty}_{r=0} e^{-\tilde{\beta} r h\nu^{s}}},\\
&\frac{E^{s}}{A^{s}}=\frac{h\nu^{s}}{e^{\tilde{\beta} h\nu^{s} }-1}.
\end{split}
\label{Eq11}
\end{eqnarray}
Using Eqns \eqref{Eq8}, \eqref{Eq10}, and \eqref{Eq11}, we can calculate the equilibrium values of energy and entropy for the entire system of photons, which can be expressed as \cite{Bose24,Bose76,Wali}
\begin{eqnarray}
\begin{split}
&\mathcal{E}= \sum_{s=0}^{\infty}E^{s}=\sum_{s=0}^{\infty}\frac{A^{s}h\nu^{s}e^{-\tilde{\beta}h\nu^{s}}}{1-e^{-\tilde{\beta}h\nu^{s}}},\\  
&\mathcal{S}_{c} = k[\tilde{\beta} \mathcal{E}+\sum_{s=0}^{\infty} A^{s} \log \mathcal{Z}], \\  
& = k[\tilde{\beta} \mathcal{E}-\sum_{s=0}^{\infty} A^{s}\log(1- e^{-\tilde{\beta} h\nu^{s}})],
\end{split}
\label{Eq12}
\end{eqnarray}
where we have used the partition function $\mathcal{Z}= \prod_{s=0}^{\infty}[Z^{s}]^{A^{s}}$ of the entire system. All results presented here pertain to a unit volume. Additionally, volume $V$ is included in these calculations through the partition function's dependence on volume, reinforcing the expression $\mathcal{Z}= \prod_{s=0}^{\infty}[Z^{s}]^{V A^{s}}$ \cite{khinchin13}. Based on quantum mechanical principles, Bose \cite{Bose24,Bose76,Wali} famously derived the expression $A^{s}= 8\pi\nu^{2}d\nu^{s}/c^{3}$. Here, $c$ represents the velocity of photons. If we substitute $A^{s}$ into Eq.\eqref{Eq12}, we obtain the results for energy and entropy as functions of volume \cite{Bose24,Bose76,Wali}.

\section{Case of Planck's Oscillators}

Planck \cite{Planck91,Planck00,Planck01} studied a system comprising a large number of identical linear oscillators, which we refer to as Planck's oscillators. He determined how these oscillators are distributed across various infinitesimal space elements in the state space of the entire system \cite{Planck91,Planck00,Planck01}. For a given thermodynamic state, Planck \cite{Planck91,Planck00,Planck01} derived expressions for the thermodynamic entropy and energy of the system as a whole. We utilize Planck's notation to examine the large deviations in the probability distributions of the thermodynamic variables associated with the oscillators.

Let $\mathbb{N}=N_{1}+N_{2}+...+N_{n}+...$ and $\mathbb{E}=E_{1}+E_{2}+...+E_{n}+...$ represent the total number of oscillators and their corresponding energies across various spatial elements $1, 2, 3, ...$ of the state space of the entire system, respectively \cite{Planck91,Planck00,Planck01}.  
The energy $E_{n}$ of the $N_{n}$ oscillators in the $n$-th space element can be expressed as $E_{n}=N_{n}(n-1/2)h\nu =  \mathbb{N}\omega_{n}(n-1/2)h\nu$, where $h$ is a constant, $h\nu/2 $ is the zero-point energy, and $\omega_{n} = N_{n}/ \mathbb{N}$ denotes the distribution density in the $n$-th space element \cite{Planck91}. Consequently, the total energy of the system across all space elements is represented as $\mathbb{E} = \mathbb{N}h\nu \sum_{n=1}^{\infty} (n-1/2) \omega_{n}$ \cite{Planck91}. Furthermore, Planck \cite{Planck91} assumed that the oscillators are uniformly distributed across the space elements.

Considering the constraints on the total number $\mathbb{N}$ of oscillators and the overall energy $\mathbb{E}$ of the system, we derive Boltzmann's distribution $\tilde{\mathcal{W}}_{b}$ for the spatial arrangements of the oscillators across all elements \cite{Planck91}.
\begin{equation}
\tilde{\mathcal{W}}_{b}\left(N_{1},N_{2},...,N_{n},...\right)= \prod \limits^{\infty}_{n=1} \frac{\mathbb{N}!}{N_{n}!}
\label{Eq13}
\end{equation}
We begin by taking the natural logarithm of both sides of the \eqref{Eq13}. Next, we apply Stirling's approximation for large numbers to the factorials of $\mathbb{N}! $ and $N_{n}!$, and then divide both sides by $\mathbb{N}$. 
\begin{eqnarray}
&& \log \tilde{\mathcal{W}}_{b}\left(.\right) = \log \mathbb{N}!-\sum^{\infty}_{n=1}\log N_{n}!, \label{Eq14} \\ 
&& \log \tilde{\mathcal{W}}_{b}\left(.\right)\approx \mathbb{N}\log \mathbb{N}-\mathbb{N} -\sum^{\infty}_{n=1}(N_{n}\log N_{n}-N_{n}), \nonumber \\ 
&& \frac{1}{\mathbb{N}} \log\tilde{\mathcal{W}}_{b}\left(.\right) \approx \sum^{\infty}_{n=1}\frac{N_{n}}{\mathbb{N}}\log \frac{\mathbb{N}}{N_{n}} = -\sum^{\infty}_{n=1}\omega_{n} \log \omega_{n}. \nonumber
\end{eqnarray}
From Eq. \eqref{Eq14}, we find that the Boltzmann-Sanov relative entropy \cite{Varadhan84,Hollander00,Ellis06,Kobayashi11} for the entire system is equal to $k$ times the logarithm of $\tilde{\mathcal{W}}_{b}$. Furthermore, the large deviation function is expressed as $ \tilde{\mathbb{I}}_{b}=-\tilde{\mathbb{S}}_{b}/k$.
\begin{eqnarray}
&& \tilde{\mathbb{S}}_{b}= k\log\tilde{\mathcal{W}_{b}}^{}\left(.\right)= -k \mathbb{N} \sum^{\infty}_{n=1}\omega_{n} \log \omega_{n}, \nonumber \\
&& \tilde{\mathbb{I}}_{b}= -\tilde{\mathbb{S}}_{b}/k = \mathbb{N}\sum^{\infty}_{n=1}\omega_{n} \log \omega_{n}. 
\label{Eq15}
\end{eqnarray}
Next, we estimate the large deviation probability for the energy of a system consisting of Planck's oscillators. This energy is the sum of the energies of the oscillators across all space elements. Specifically, we use the BCM inequality to compute the Cramér-Chernoff large deviation upper bound on the probability of the exponential random variable $e^{-\tilde{\beta}\mathbb{E}}$, where $\tilde{\beta}={(kT)}^{-1}>0$ \cite{Varadhan84,Hollander00,Ellis06,Kobayashi11}. Consequently, for a given specific energy threshold $\mathbb{N}u$, we establish a large deviation  upper bound of the probability inequality.
\begin{eqnarray}
&& \tilde{\mathcal{P}}_{c}(e^{-\tilde{\beta} \mathbb{E}} \geq e^{-\tilde{\beta} \mathbb{N}u}) \leq \frac{\mathbb{E}_{x}(e^{-\tilde{\beta} \mathbb{E}})}{e^{-\tilde{\beta} \mathbb{N}u}} = \frac{(Z_{1})^{\mathbb{N}}}{e^{-\tilde{\beta} \mathbb{N}u}} \label{Eq16}, \\ \nonumber 
&& \log \tilde{\mathcal{P}}_{c}(.) \leq \tilde{\beta} \mathbb{N}u + \mathbb{N} \log Z_{1}. \\ 
&& \frac{1}{\mathbb{N}}\log \tilde{\mathcal{P}}_{c}(.)\leq \tilde{\beta}u + \log Z_{1}. \label{Eq17}
\end{eqnarray}
In Eq. \eqref{Eq16}, we use the partition functions $Z_{1}(\tilde{\beta})$ and $\mathbb{Z}(\tilde{\beta})=Z_{1}^{\mathbb{N}}$ for an oscillator and the entire system, respectively. The lower line of Eq. \eqref{Eq16} presents the logarithm of probability, which describes the Boltzmann-Gibbs canonical entropy $\tilde{\mathcal{S}}_{c}$ \cite{LandauV5,Schroedinger13,Haar95} associated with a specific macrostate. As the limit $\mathbb{N}$ approaches infinity, Eq. \eqref{Eq17} yields an expression for the Cramer-Chernoff large deviation function $\tilde{\eta}_{c}$.
\begin{eqnarray}
&& \tilde{\mathcal{S}}_{c}= \min_{\tilde{\beta}}\left[k\tilde{\beta} \mathbb{N}u+ k \mathbb{N} \log Z_{1} \right] \label{Eq18}. \\ 
&& \tilde{\eta_{c}}= - \frac{\tilde{\mathcal{S}_{c}}}{k} = \max_{\tilde{\beta}}\left[-\tilde{\beta} \mathbb{N}u - \mathbb{N} \log Z_{1} \right]  \nonumber
\end{eqnarray}
The condition for thermodynamic equilibrium requires that the $\delta$ variations of the entropies or the logarithms of the probability distributions with respect to $N_{n}$ be equal to zero \cite{Planck91}. We can determine the equilibrium probability \cite{Planck91} by taking the $\delta$-variations of Eqns \eqref{Eq15} and \eqref{Eq16}.
\begin{eqnarray}
\begin{split}
& \sum^{\infty}_{n=1}\left[\log \frac{\mathbb{N}}{N_{n}}- \tilde{\beta} (n-1/2) h \nu -\log Z_{1}\right] \delta N_{n} =0,\\
&\frac{N_{n}}{\mathbb{N}}= \omega_{n}=\frac{e^{-\tilde{\beta} (n-1/2) h \nu}}{Z_{1}(\tilde{\beta})}. 
\end{split} 
\label{Eq19}
\end{eqnarray}
If we substitute the lower line of Eq. \eqref{Eq19} into the condition $\mathbb{N}=\sum_{n}N_{n}$, we derive the expression for the partition function $Z_{1}=\sum_{n}e^{-\tilde{\beta} (n-1/2)h\nu}$ of an oscillator. The partition functions of an oscillator and the system are used to calculate the average energies, $\overline{u}$ and $\mathbb{E}$ for the oscillator and the entire system respectively \cite{Planck91}.
\begin{eqnarray}
\begin{split}
& \overline{u}= -\frac{\partial \log Z_{1}}{\partial \tilde{\beta}}= \frac{h\nu}{2}+\frac{h\nu}{e^{\tilde{\beta}h\nu}-1},\\
&\mathbb{E}=\mathbb{N}\overline{u}= N\left(\frac{h\nu}{2}+\frac{h\nu}{e^{\tilde{\beta}h\nu}-1}\right). 
\end{split} 
\label{Eq20}
\end{eqnarray}
Additionally, the entropy of the entire system of Planck's oscillators \cite{Planck91} is calculated as follows:
\begin{eqnarray}
\begin{split}
& \tilde{\mathcal{S}}_{c}= k \mathbb{N} \left[\tilde{\beta} \overline{u}+\log Z_{1}\right],\\
& \tilde{\mathcal{S}}_{c}= k \mathbb{N} \left[ \frac{\tilde{\beta}h\nu}{e^{\tilde{\beta}h\nu}-1} - \log\left(1-e^{-\tilde{\beta}h\nu} \right) \right]. 
\end{split} 
\label{Eq21}
\end{eqnarray}
The results of Planck's oscillators are similar to those of one-dimensional quantum mechanical oscillators \cite{LandauV5,Schroedinger13,Haar95,Pathria11,Greiner95}.
\section{Summary}
In this article, we demonstrate the applications of the large deviation theorem to the distributions of a photon gas and Planck's oscillators for a given macroscopic state. Specifically, we utilize the Cramér-Chernoff theorem to calculate the upper bound for the large deviation probability of energy random variables. Furthermore, we have reproduced the equilibrium values of probabilities, thermodynamic entropies, energies, and occupation numbers in both cases.

\bibliography{GrandRef}
\end{document}